\documentclass[aps,pre,floatfix,superscriptaddress,amsmath,showpacs,twocolumn]{revtex4}
\usepackage{graphicx}
\newcommand {\be}{\begin{equation}}
\newcommand {\ee}{\end{equation}}

\begin{document}

\title{Multiplicative noise in the longitudinal mode dynamics of a
bulk semiconductor laser}

\author{Francesco Pedaci}
\affiliation{Institut Non-lin\'{e}aire de Nice, UMR 6618 Centre
National de la Recherche Scientifique - Universit\'{e} de Nice
Sophia-Antipolis, 06560 Valbonne, France}

\author{Stefano Lepri}
\affiliation{Istituto dei Sistemi Complessi, Consiglio Nazionale
delle Ricerche, via Madonna del Piano 10, 50019 Sesto Fiorentino, Italy}

\author{Salvador Balle}
\affiliation{Institut Mediterrani d'Estudis Avan\c cats, IMEDEA
(CSIC-UIB), C/ Miquel Marqu\'es 21, E-07190 Esporles, Spain}

\author{Giovanni Giacomelli}
\affiliation{Istituto dei Sistemi Complessi, Consiglio Nazionale
delle Ricerche, via Madonna del Piano 10, 50019 Sesto Fiorentino, Italy}

\author{Massimo Giudici} \affiliation{Institut
Non-lin\'{e}aire de Nice, UMR 6618 Centre National de la Recherche
Scientifique - Universit\'{e} de Nice Sophia-Antipolis, 06560
Valbonne, France}

\author{Jorge R. Tredicce}
\affiliation{Institut Non-lin\'{e}aire de Nice, UMR 6618 Centre
National de la Recherche Scientifique - Universit\'{e} de Nice
Sophia-Antipolis, 06560 Valbonne, France}

\date{\today}

\begin{abstract}

We analyze theoretically and experimentally the influence of current noise
on the longitudinal mode hopping dynamics of a bulk semiconductor laser.
It is shown that the mean residence times on each mode have different
sensitivity to external noise added to the bias current. In particular, an
increase of the noise level enhances the residence time on the
longitudinal mode that dominates at low current, evidencing  the
multiplicative nature of the stochastic process.  A two-mode rate equation
model for semiconductor laser is able to reproduce the experimental
findings. Under a suitable separation of the involved time scales, the
model can be reduced to a 1D bistable potential system with a
multiplicative stochastic term related to the current noise strength. The
reduced model clarifies the influence of the different noise sources on
the hopping dynamics.

\end{abstract}
\maketitle

\section{Introduction}

Fluctuations and noise are inherent to the behavior of any
physical system. Their ubiquity and impact on technological
applications have stimulated since long ago the study of their
effects in different branches of science (statistical physics,
mathematics) and technology (communications engineering).

Noise is usually perceived as a source of degradation of the
properties of any system, leading to relatively small stochastic
variations of a given magnitude around its deterministic (noise
free) value. Such a point of view is rooted on the study of systems
close to equilibrium, whose dynamics can be linearized around the
deterministic steady state. However, it does not hold in systems far
from equilibrium, where nonlinear corrections to the dynamics must
be taken into account. In these systems, noise may not only be the
responsible for the observed behavior, but it may also make possible
new states and behaviors that do not exist in the noise-free limit
\cite{toral}. Examples of such behaviors are, for instance, the
enhancement of the decay time of a metastable state (noise enhanced
stability) \cite{Graham,Mantegna}, the synchronization with a weak
periodic input signal (stochastic resonance) \cite{sto_res} or the
appearance of a periodic output (coherence resonance)
\cite{coh_res}. In systems with spatial degrees of freedom, noise may
lead for instance to the formation of convective structures
(noise-sustained structures) \cite{maxi}.

The effects of noise depend on whether it is additive or
multiplicative. Additive noise is present in all real, non
isolated systems where the environment acts as a thermal bath.
Moreover, even in an isolated system all interaction processes
exhibit some degree of stochastic fluctuations that lead to
internal noise in the system. This effect may be further enhanced
by adding noise to the control parameters of the system. In
general, the fluctuation-dissipation theorem implies that noise in
a nonlinear system may have both an additive and a multiplicative
component. The effects of both kinds of noise in nonlinear systems
have been thoroughly studied from the theoretical point of view
\cite{toral}, although experimental studies of the effects of
multiplicative noise are more scarce. In particular, it has been
shown that the characteristic signatures of either type of noise
can be profoundly modified by the presence of even a weak
component of the other kind \cite{Schenzle,Mannella}. Moreover it
was also found that multiplicative (parametric) noise induces a
shift in the critical mean value of the parameter that controls an
instability \cite{Lefever,Moss,Fronzoni}.

In this paper we analyze theoretically and experimentally the
influence of multiplicative noise on the mode hopping dynamics of a
bistable semiconductor laser. Since the pumping current enters into
the laser field equations in a nonlinear way, adding current noise
may lead to multiplicative effects. We find that the residence times
of each mode are strongly affected, and that increasing the noise
added to the current is equivalent ---from the point of view of
residence times--- to a lowering of the bias current. A theoretical
model for the system under study is presented and analyzed, with the
aim of understanding the observable consequences of the imposed
fluctuations and the role of laser parameters. A better insight is
obtained by simplification of this model to a bistable potential
system. This allows us to  discuss in detail the effects of current
noise on the mode hopping dynamics. The results are in good
agreement with the experimental observations carried out for a bulk
edge-emitting semiconductor laser.

The outline of the paper is the following. In Sec.~II we introduce
the rate equations for a two-mode semiconductor laser and we derive
from them a reduced 1D Langevin model that describes the hopping
dynamics of the freely operating laser. In Sec.~III we discuss in
detail the reduced model and, in particular, the effect of external
fluctuations of the injection current. The limit case of a slowly
fluctuating current is analyzed as well and the results are compared
with simulation of the rate equations. Sec.~IV is devoted to the
presentation of the experimental results. We draw our conclusions in
Sec.~V.

\section{Theoretical analysis}

\subsection{Rate equations}

Our starting point is a stochastic rate-equation model for a
semiconductor laser that may operate in two modes. Both modes
interact with a single carrier density that provides the necessary
gain. The two modes have very similar linear gains, provided that
their wavelengths are almost equal and they are close to the gain
peak. If $E_\pm$ denote the complex modal amplitudes, $N$ the
carrier density and $J(t)$ the injection current, the model can be
written as
\begin{eqnarray}
&\dot E_+& = \frac{1}{2}\left[ (1 + i \alpha) g_+ - 1 \right] E_+
+ \sqrt{2D_{sp} N}\, \xi_+ \label{rateq1} \\
&\dot E_-& = \frac{1}{2}\left[ (1 + i \alpha) g_- - 1 \right] E_-
+ \sqrt{2D_{sp} N}\, \xi_- \label{rateq2} \\
&\dot N  & = \gamma[ J(t) - N - g_+ |E_+ |^2 - g_- |E_- |^2]
\label{rateq3}
\end{eqnarray}
where $\alpha$ is the linewidth enhancement factor
\cite{petermann}. The modal gains read
\begin{equation}
g_\pm \;=\; {N \pm \varepsilon(N - N_c) \over 1+ s |E_\pm |^2 +c
|E_\mp |^2} \; ,
\end{equation}
where $\varepsilon$ determines the difference in differential gain
among the two modes while $N_c$ defines the carrier density where
the unsaturated modal gains are equal. The parameters $s$ and $c$
are respectively the self- and cross-saturation coefficients. The
$\xi_\pm$ are two independent, complex white noise processes with
zero mean ($\langle\xi_{\pm}(t)\rangle = 0$) and unit variance
($\langle\xi_i (t) \xi_j^*(t')\rangle = \delta_{ij} \delta(t-t')$)
that model spontaneous emission. Eqs.~(\ref{rateq1}-\ref{rateq3})
have to be interpreted in It\^o sense \cite{Agrawal}, thus the
average power spontaneously emitted in each mode at any time is
given by $4 D_{sp} N$.

The deterministic version of Eqs.~(\ref{rateq1}-\ref{rateq3})
admits four different steady state solutions: the trivial solution
$E_{\pm} = 0$, two single-mode solutions --- $E_+ \neq 0 , \ E_- =
0$ and viceversa --- and a solution where both modes are lasing,
$E_{\pm} \neq 0$. The single-mode solution where only $E_+$
($E_-$) is lasing lacks physical sense for bias currents below
$J_+$ ($J_-$) given by
\begin{equation}
J_{\pm} = {1 \pm \varepsilon N_c \over 1 \pm \varepsilon} \; .
\end{equation}
The trivial solution $E_{\pm} = 0$ is the only stable solution for
bias currents $J < min(J_+, J_-)$; for definiteness,
we consider $N_c > 1$, hence $J_- < 1 < J_+$. Upon increasing $J$,
the trivial solution looses stability and the system switches to the
solution $E_- \neq 0, \ E_+ = 0$ at the laser threshold, $J = J_-$.
Further increasing the current, the sequence of bifurcations depends
quite strongly on the parameters $s$, $c$ and $N_c$. For $c > s$,
this solution may coexist with the solution $E_+ \neq 0, \ E_- = 0$
or even with the solution $E_- \neq 0, \ E_+ \neq 0$ (in this case,
only for a small current range). Finally, the solution $E_+ \neq 0,
\ E_- = 0$ prevails.

\subsection{Reduction to an effective model}

In order to better assess the effects of current noise on the modal
dynamics, we next reduce the rate-equation description to
a bistable 1D system. In the first place, we introduce
the amplitude-phase coordinates for each mode,
\begin{equation}
E_{\pm} \;=\; \rho_{\pm} \exp i\psi_{\pm}.
\end{equation}
Using this standard transformations (see \cite{gard}) we
obtain
\begin{eqnarray}
\dot \rho_{\pm} &=& \frac{1}{2} \Big[ g_{\pm} -1 + \frac {2D_{sp} N} {\rho_{\pm}^2} \Big] \rho_{\pm}
   + \sqrt{2D_{sp} N}\, \xi_\rho^{\pm} \nonumber \\
\dot \psi_{\pm} &=& \frac{\alpha}{2} g_{\pm} + \sqrt{2D_{sp} N}\, \xi_\psi^{\pm}\\
\dot N &=& \gamma[ J(t) - N - g_+ \rho_+^2 - g_- \rho_-^2 ] \nonumber \; .
\end{eqnarray}
Since the modal phases do not influence the evolution of the modal
amplitudes and carrier density, we can disregard them without loss
of generality. It is convenient to perform a further change
to ``cylindrical'' coordinates,
\begin{equation}
\rho_+ = r \cos \phi ,\qquad  \rho_- = r \sin \phi \; .
\end{equation}
In these new variables, $r^2$ is the total power emitted by
the laser, and $\phi$ determines how this power is partitioned
among the two modes: $\phi=0$ ($\phi = \pi/2$) corresponds to
emission in mode + (-) only, and intermediate values give
different power to each mode.

Using again Ref.~\cite{gard}, we obtain
\begin{widetext}
\begin{eqnarray}
\dot r &=& \frac {r}{2} \left[
   \frac {N \left( 1 + \sigma r^2 + \delta r^2 \cos^2 2\phi \right) +
          \varepsilon \left( N - N_c \right) \left( 1 + \sigma r^2 + \delta r^2 \right) \cos 2\phi }
         {\left( 1 + \sigma r^2 \right)^2 - \left( \delta r^2 \cos 2\phi \right)^2}
   -1 + \frac{6 D_{sp} N}{r^2} \right] + \sqrt{ 2 D_{sp} N } \xi_r \; , \\
\dot \phi &=& -\frac{ \sin 2 \phi} {2}
   \frac {N \delta r^2 \cos 2\phi + \varepsilon \left( N - N_c \right) \left( 1 + \sigma r^2 \right) }
         {\left( 1 + \sigma r^2 \right)^2 - \left( \delta r^2 \cos 2\phi \right)^2}
   + \frac {2 D_{sp} N} {r^2 \tan 2\phi} + \sqrt{ \frac{2 D_{sp} N }{r^2}} \xi_{\phi} \; , \\
\dot N &=& \gamma \left[ J - N - r^2 \frac {N \left( 1 + \sigma r^2 + \delta r^2 \cos^2 2\phi \right) +
          \varepsilon \left( N - N_c \right) \left( 1 + \sigma r^2 + \delta r^2 \right) \cos 2\phi }
         {\left( 1 + \sigma r^2 \right)^2 - \left( \delta r^2 \cos 2\phi \right)^2} \right] \; .
\end{eqnarray}
\end{widetext}
where we have defined
\begin{equation}
\sigma = {c + s \over 2}, \quad \delta = {c - s \over 2} .
\end{equation}
The parameter $\sigma$ thus represents the gain saturation induced by the
total power in the laser, while $\delta$ ---given by the
difference between the coefficients of cross- and
self-saturation--- describes the reduction in gain saturation due
to partitioning of the power among the two modes. We also wish to
remark that, although we worked in the It\^o formalism, this form
of the equations would be the same also in the Stratonovich
representations \cite{gard}.

The general expression just obtained for the dynamics is too
involved to allow for a detailed analysis of the effects of noise.
In order to simplify the theoretical analysis of this dynamical
system, we further assume that: (i) The asymmetry of the modal gains
is very small, i. e., $N_c \gtrsim 1$, $\varepsilon \ll 1$,
$\delta \ll 1 $; (ii) the laser operates close to threshold, so that the saturation
term is small, $\sigma r^2 \ll 1$. In this limit, we find that, to lowest
order in the small terms, the dynamics is governed by
\begin{widetext}
\begin{eqnarray}
\dot r &=& \frac {r}{2} \left( N - 1 - N \sigma r^2 + \frac{6
D_{sp} N}{r^2} \right) + \sqrt{ 2 D_{sp} N } \xi_r \; ,
\label{rdot} \\
\dot \phi &=& -\frac{ \sin 2 \phi} {2}
   \left[ N \delta r^2 \cos 2\phi + \varepsilon \left( N - N_c \right) \right]
   + \frac {2 D_{sp} N} {r^2 \tan 2\phi} + \sqrt{ \frac{2 D_{sp} N }{r^2}} \xi_{\phi} \; ,
\label{phidot1} \\
\dot N &=& \gamma \left( J - N - r^2 N \right) \; , \label{Ndot}
\end{eqnarray}
\end{widetext}
so the dynamics of the total intensity $r^2$ and of the
carrier density $N$ decouple from those of the relative phase
$\phi$. Thus $\phi$ is driven by the other two variables.

Moreover, the time-scale for the evolution of the relative phase
$\phi$ is of second order in the small quantities, while that of
$r$ and $N$ is of the first order. Since we are mainly interested
in time scales long enough for the relaxation oscillations of
the total power and carrier density to be totally damped, and not
in the transient dynamics, we can consider that $r$ and $N$ have
reached the vicinity of their steady state. By neglecting the
spontaneous emission term in (\ref{rdot}), we have that
\begin{equation} r \simeq r_0 = \sqrt{J-1 \over 1+\sigma J}; \quad  N
\simeq N_0 = {1+\sigma J \over 1+\sigma} \; . \label{r0}
\end{equation}
A more accurate approximation would be to solve the Langevin
equations for $r$ and $N$ and to insert the solution into the
equation for $\phi$ in order to describe how the fluctuations of the
former affect the dynamics of the latter. For simplicity, we
disregard this issue under the assumption that the noise in
(\ref{rdot}) and (\ref{Ndot}) is so weak that they simply "follow"
$J$ as prescribed by (\ref{r0}). For a time-dependent current this
approximation is valid only if $J$ does not change too fast.
For example, in the case of an Orstein--Uhlenbeck process
that will be considered below one must require the corresponding
correlation time $\tau$ to be longer than the typical relaxation
time of the total intensity. We shall see that this condition is
generally met in our experimental setups. 

In this representation the dynamics can be geometrically visualized
as follows. The motion is constrained along a manifold (approximated
as portion of a circle of radius $r_0$) connecting the fixed points.
Radial fluctuations (i.e. fluctuations in the total
intensity output) are completely neglected. The hopping dynamics is
thus effectively one-dimensional and described by the "phase"
variable $\phi$, which determines how the total power is partitioned
among the modes.

Altogether, the above calculation yields the reduced model
\begin{equation}
\dot \phi\; =\;
-\frac{1}{2}\Big[a \cos2\phi + b\Big]\sin2\phi \,+\, {2D_\phi\over \tan 2\phi}
 + \sqrt{2 D_\phi}\, \xi_\phi
\label{phidot}
\end{equation}
where we have defined the parameters
\begin{eqnarray}
&  a \;=\;& N_0\delta \, r_0^2    \;=\; {\delta  \over 1+\sigma}\,(J-1) \\
&  b\;=\;&\varepsilon(N_0-N_c) \;=\; {\varepsilon \sigma \over 1+\sigma}\,
(J-J_s) \quad \\
&  D_\phi \;=\;&{D_{sp} N_0 \over r_0^2} \;=\;
\frac {( 1 + \sigma J)^2} {(1 + \sigma) (J-1)} \, D_{sp}
\label{params}
\end{eqnarray}
and, for later convenience, we have introduced
\begin{equation}
J_s \;=\; {(1+\sigma) N_c -1 \over \sigma}\quad.
\label{Js}
\end{equation}
Notice that the strength of fluctuations $D_\phi$ depends on the
current. At this stage, one may get rid of one of the three
parameters. For example, one may rescale time $t \to t/a; \xi_\phi
\to \sqrt{a} \xi_\phi$ so that the only independent parameters are
$b/a$ and $D_\phi/a$. However, since we wish to understand
the role of the phenomenological parameters in relation with the
physical quantities we stick to the form (\ref{phidot}). This choice
is especially useful when current fluctuations will be taken into
account.

To conclude this Section, we point out that the same equation
(\ref{phidot}) has been derived by Willemsen et
al.~\cite{will,will2} to describe polarization switches in Vertical
Cavity Surface-Emitting Lasers (VCSELs). The starting point of their
derivation is the San Miguel-Feng-Moloney model~\cite{Feng}. We
point out that the physical meaning of the variable $\phi$ is
different from here as it represents the polarization angle of
emitted light. Thus the potential minima correspond to the two
orthogonal linearly--polarized directions. Also, Nagler et
al.~\cite{belgi,belgi2} have derived a one-dimensional Langevin
equation for the mode intensity in VCSEL starting from a rate
equation model for the field intensities in the two polarization
directions. Indeed, it can be shown that, upon the change of
variable $\sin\phi = p$ and redefinition of parameters,  model
(\ref{phidot}) transforms into eq. (19) and (23) of
Ref.~\cite{belgi}. This suggests that, upon a suitable
reinterpretation of variables and parameters, many of the results
presented henceforth may apply also to the dynamics of VCSELs.

\section{Analysis of the reduced model}

In this Section we analyze some features of the model in the case in  which $J$
is a constant. In this case, (\ref{phidot}) is a one-dimensional Langevin
equation with additive noise originating from spontaneous emission. The force
in Eq.~(\ref{phidot}) can be derived from the potential
\begin{equation}
U(\phi,D_\phi) \;=\; -{a\over 16} \cos 4\phi - {b\over 4}\cos 2\phi -
D_\phi \ln \sin 2\phi
\label{uphi}
\end{equation}
The stationary distribution is thus straightforwardly computed as
\begin{equation}
P(\phi) \;=\; P_0 \, \exp \left( -U(\phi,D_\phi)/D_\phi \right),
\label{ps}
\end{equation}
with $P_0$ being the normalization constant. Notice that the
potential depends explicitly on $D_\phi$ (i.e. on $D_{sp}$) through
the repulsive logarithmic term in (\ref{uphi}). This term is usually
very small for weak noise except at the extrema $\phi=0, \pi/2$
where it diverges logarithmically. As a consequence, $P$ vanishes
linearly there. Physically, this corresponds to the fact that, due
to spontaneous emission, the modes are never completely switched
off.

\subsection{Bistability and effect of the injection current}
\label{s:bist}

The most important property that can be immediately drawn from the
form  of $U$ is that there exists a range of current values $J_3
<J<J_4 $ for which the system is bistable. Within this region, the
system has three stationary solutions, $\phi_\pm$ (stable) and
$\phi_0$ (unstable). To compute them, let us neglect the term
proportional to $D_\phi$ in the definition of $U$: $\phi_0\simeq
(1/2) {\rm acos}(-b/a)$, $\phi_-\simeq 0$, $\phi_+\simeq \pi/2$. The
bistability domain is evaluated by the condition $|b|\le a$ yielding
the approximate expressions
\begin{equation}
J_3 \;=\;
{\varepsilon \sigma J_s + \delta \over \varepsilon \sigma+ \delta },\qquad
J_4 \;=\;
\begin{cases}
{\textstyle\varepsilon \sigma J_s -
\delta \over \textstyle\varepsilon \sigma - \delta}
& \text{if $\varepsilon \sigma > \delta$,}\\
+\infty &\text{otherwise.}
\end{cases}
\end{equation}
Those last expressions have been checked to be a good approximation
of the exact values computed from bifurcation analysis of the
deterministic rate equations in the limit of small $\varepsilon$ and
$\delta$. The effect of the term proportional to $D_\phi$ in
(\ref{uphi}) is to make the bistability domain dependent on
the noise intensity, i.e. $J_3(D_\phi)<J<J_4(D_\phi)$. By computing
numerically the stationary points  of $U(\phi,D_\phi)$ we have found
that the width of this interval of current values is reduced upon
increasing $D_\phi$ with respect to the deterministic case.
The current value $J$ controls the symmetry of the potential through
the term proportional to $b$. Within the bistable region, at the
critical value $J_s$, $b$ vanishes and the effective potential is
symmetric under the transformation $\phi\to \pi/2 -\phi$. In this
situation the hopping between the two modes occurs at the same rate.
For weak noise, $D_\phi \ll a $, we can easily estimate the two
potential barriers as
\begin{equation}
\Delta U_\pm \;=\; \frac{(a \pm b)^2}{8a} \quad .
\label{barri}
\end{equation}
In the neighbourhood of the  $J\simeq J_s$, i.e. for $(J-J_s) \ll (J_s-1)$:
\begin{equation}
\Delta U_\pm \;\simeq\; \frac{\delta}{8(1+\sigma)}(J_s-1)
\,+\, \frac{\delta \pm 2\varepsilon\sigma}{8(1+\sigma)}
(J -J_s) \; .
\label{barris}
\end{equation}
Actually, also the noise strength depends on $J$: expanding to second order
around $J_s$ the definition of $D_\phi$, Eq.~(\ref{params}), we get
\begin{eqnarray}
D_\phi && \;\simeq\; D_\phi(J_s)
\Big[1 - {\sigma(J_s-2) -1\over (1+\sigma J_s)(J_s-1)}(J -J_s) + \nonumber\\
&&{(1+\sigma)^2\over (1+\sigma J_s)^2(J_s-1)^2}(J -J_s)^2 \Big]
\label{line}
\end{eqnarray}
However, it turns out that this correction to the noise is pretty
small for the parameters we choose. More precisely, one can estimate
that, for a given change of $J$, the relative change
of $\Delta U_\pm$ is roughly a factor 10 larger than the one of
$D_\phi$ \cite{note}. We thus neglect the correction terms in
(\ref{line}) and let $D_\phi=D_\phi(J_s)$.

In other words, we assume that changing $J$ only affects the
barriers through formula (\ref{barris}). With this simplification
the corresponding residence times are given by \cite{rmp}
\begin{eqnarray}
T_\pm &\;=\;& \sqrt{8\pi D_\phi \over a^3}\, \exp\Big(\Delta U_\pm/D_\phi\Big),
\nonumber \\
&\;\equiv\;& T_s
\exp\Big[\frac{\delta \pm 2\varepsilon \sigma}{8(1+\sigma)}\,
\frac{J -J_s}{D_\phi}
\Big]
\label{tpm}
\end{eqnarray}
Here $T_s$ is the residence time at the symmetry point.
Notice that the prefactor depends on the noise strength to leading
order. Formula (\ref{tpm}) enlightens the role of the asymmetry
parameters in determining the residence times. Indeed, it
predicts that the residence times should depend
exponentially on the injection current with different rates. In
particular, we see that increasing $J$ around $J_s$ may lead to an
increase of both $T_\pm$ if $2\varepsilon\sigma<\delta$ or to an
increase of $T_+$ accompained by a decrease of $T_-$ if instead
$2\varepsilon\sigma>\delta$. To discriminate which of the two cases
is of relevance one must thus know the values of the parameters for
the laser at hand. This can only be accomplished by comparing with
experimental data. For the laser at hand (see below), the
measurements indicate that the case of interest is
$2\varepsilon\sigma>\delta$.

\subsection{Comparison with the simulation of the rate equations}
\label{s:simu}

Before proceeding further, we check the accuracy of the reduction by
comparing with simulations of the rate equations. For defineteness,
we let $\varepsilon=0.1$, $s=1.0$, $N_c=1.1$, $\gamma=0.01$ and $c=
1.1$ and perform different runs for different values of $D_{sp}$ and
for $J$ some  10\% above threshold. With the above choice,
$\delta=0.05$ and $\sigma=1.05$, meaning that the timescale for the
slow variable $\phi$ is estimated to be about one order of magnitude
larger than those of the fast ones. We also set $\alpha=0$ since, as
explained above, we expect that the phase dynamics should not
matter. The largest part of the simulations were performed
with Euler method with time steps 0.01-0.02. For comparison, some
checks with Heun method \cite{toral} have also been carried on.
Within the statistical accuracy, the results are found to be
insensitive the the choice of the algorithm.

\begin{figure}[h]
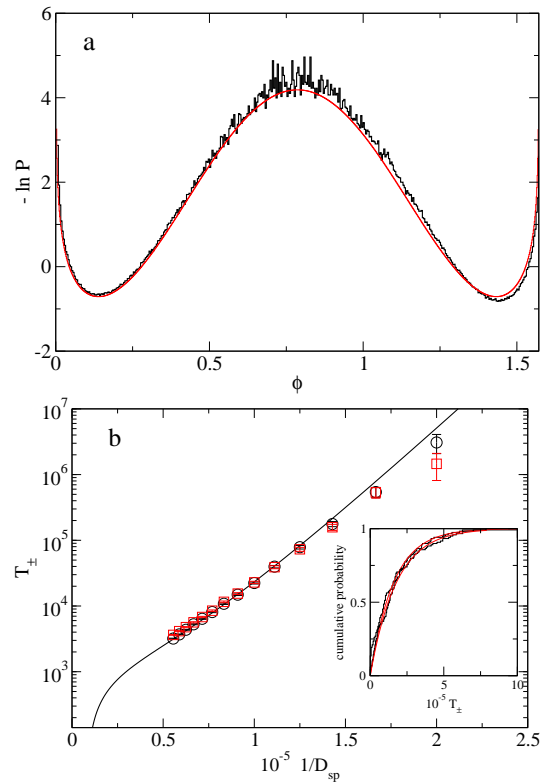

\begin{center}
\includegraphics[clip,width=7cm]{fig1a.eps}
\includegraphics[clip,width=7cm]{fig1b.eps}
\caption{
(Color online)
Comparison between the reduced model and the simulation of the rate equations
for $J=1.197$, $D_{sp}=7\times 10^{-6}$ (the other parameter values are  given in
the text). (a) $-\ln P$ evaluated from simulation compared with the
analytical result;  (b) Average residence times as a function of $1/D_{sp}$.
The line is obtained from the quadrature formula (\ref{quadra}). Inset:
Cumulative distributions of residence times for $D_{sp}=7\times 10^{-6}$. The lines
are the expected Poisson distribution with the same average.}
\label{f:comp}
\end{center}
\end{figure}

We first evaluated the stationary distribution $P(\phi)$ as evaluated from
the data through the relation $\phi = {\rm atan} (|E_+|/|E_-|)$.  In
Fig.~\ref{f:comp} we compare the expression of $U$ with $-\ln P$. The
simulation data reported there correspond to the current value $J=1.197$ which
we empirically found to yield an almost symmetric shape of $P$. This value
only differs by 2\% from the one estimated from formula (\ref{Js}) yielding
$J_s=1.19523...$.   We also compared the numerically measured residence times
$T_\pm$ with the prediction of the reduced model (see again
Fig.~\ref{f:comp}).  Since we work with values of the noise that are not
extremely small, we employed the quadrature formula \cite{rmp}
\begin{equation}
T_\pm \;=\; \pm \frac1D_\phi \int_{\phi_-}^{\phi_+}
P^{-1}(x) \int_{\phi_\pm}^{x} \, P(y) \, dxdy \quad .
\label{quadra}
\end{equation}
This functional form accurately follows the numerical data
(within then 10\% or so). In the inset of Fig.~\ref{f:comp} we show that the
cumulative distribution of residence times which is Poissonian as expected from
the reduced model.  Another series of simulations (not reported) for an
asymmetric  case, $J=1.18$ yield comparable results.

\subsection{Including current fluctuations}

We include the fluctuation of the injected current by letting $J
\longrightarrow J + \delta J(t)$. The following considerations holds for an
arbitrary time dependencence of $\delta J$ under the limitations necessary to
eliminate Eq.~(\ref{rdot}). For defineteness, we may focus on the
case in which $\delta J$ is a Ornstein-Uhlenbeck process with zero average
$\langle\delta J(t)\rangle \;=\;0$ and correlation time $\tau$:
\begin{equation}
\dot{ \delta J} \;=\; -{\delta J \over \tau} + \sqrt{2D_J\over \tau}
\,\xi_J
\label{ou}
\end{equation}
that means
\begin{equation}
\langle\delta J(t) \delta J(0)\rangle \;=\; {D_J} \exp(-|t|/\tau) \quad .
\end{equation}
This choice is suitable to model the finite-bandwith of the noise
generator employed in the experimental setup. Notice that $\tau$ and the
variance of fluctuations ${D_J}=\langle\delta J^2\rangle$ can be
fixed independentely.

The effect of a time-dependent current is twofold. First, the potential becomes
fluctuating: the $a$, $b$ and $D_\phi$ coefficients become time-dependent
quantities and the potential barriers  $\Delta U_\pm (t)$ change accordingly.
For weak noise, and close to the symmetry point
$\delta J \ll (J_s-1)$ they are computed from approximated formula
(\ref{barris}) with $J-J_s$ replaced by $\delta J(t)$,
\begin{equation}
\Delta U_\pm (t) \;\simeq\; \frac{\delta}{8(1+\sigma)}(J_s-1)
\,+\, \frac{\delta \pm 2\varepsilon\sigma}{8(1+\sigma)}
\delta J(t) \, .
\label{fbar}
\end{equation}
The barriers' height depends linearly on $\delta J$ to leading
order. Obviously, this last expression makes sense only when the
fluctuating term is sub-threshold i.e. whenever the system is
bistable (a large enough $\delta J$ could always occur making the
potential single-well). The second effect is on the additive noise
strength. Since $D_\phi$ depends on $\delta J$ the amplitude of
spontaneous emission is renormalized. Indeed, since $\xi_\phi$ is
$\delta-$~correlated, the process $\sqrt{2 D_\phi(J)}\, \xi_\phi$
can be replaced by $\sqrt{2\langle D_\phi\rangle}\, \xi_\phi$ where
the average is over the fluctuations of the variable $\delta J$.
Using the expansion (\ref{line}) we get
\begin{equation}
\langle D_\phi\rangle \;=\; D_\phi(J_s)
\Big[1 + {(1+\sigma)^2\over (1+\sigma J_s)^2(J_s-1)^2}\, D_J \Big]
\label{deff}
\end{equation}
The coefficient in front of $D_J$ is strictly positive meaning that current
fluctuations always enhance spontaneous ones. However,  as argued in Section
\ref{s:bist} for the case of static changes of $J$, it turns out that the
relative change $\Delta U_\pm$ is larger than that of $D_\phi$ if $D_J \ll
(J_s-1)^{2}$ (see however again note \cite{note}). Therefore, in the following
we can safely set  $\langle D_\phi\rangle \;\simeq\;D_{\phi}(J_s)$.

Although expression (\ref{fbar}) is sufficient to draw some conclusion
than can be experimentally tested, it is useful to write down also
the full Langevin equation associated with the problem.
Putting al the terms together we find that, to first order in $\delta J$,
equation (\ref{phidot}) transforms to
\begin{equation}
\dot \phi \;=\; - U'(\phi,D_\phi) - V'(\phi,D_\phi)\,\delta J
+ \sqrt{2 D_\phi} \, \xi_\phi
\label{langmu}
\end{equation}
where
\begin{equation}
- V'(\phi,D_\phi)\;=\; -\frac{1}{2(1+\sigma)}
  \Big( \delta \cos2\phi + \varepsilon\sigma\Big) \sin2\phi
\end{equation}
(for simplicity we neglected again the dependence of $D_\phi$ on the
current fluctuations). The multiplicative term can be thus derived
from the ``potential"
\begin{equation}
V(\phi) \;=\; -{\delta\over 16(1+\sigma)} \cos 4\phi -
{\varepsilon\sigma\over 4(1+\sigma)}\cos 2\phi .
\end{equation}
For an arbitrary choice of the parameters, $V$ has a different
symmetry with respect to $U$ meaning that the effective amplitude of
multiplicative noise is different within the two potential wells. If
this difference is large enough, current fluctuation will remove the
degeneracy between the two stationary solutions.

Non-Markovian equations of the form (\ref{langmu}) have been
thoroughly studied in the literature (see e.g.
\cite{h94,h95,iw96,m96} and  references therein). Although their
full analytical  solution for arbitrary $\tau$ is not generally
feasible, several approximate results can be provided in some
limits. In the following Section we will discuss the case which is
of experimental importance.

\subsection{The kinetic limit}

Altogether, the mode switching can be seen as an activated escape
over fluctuating barriers given by (\ref{fbar}). The statistical
properties of the latter process is controlled by the current
fluctuations. To assess the nature of the stochastic process at
hand, it is important to discuss the relevant time scales. In
particular, one should compare the relaxation time $T_R$ within the
wells with both $\tau$ and the residence times $T_\pm$. If
$T_R < \tau$ we are in the colored noise case. An estimate of $T_R$
is the inverse of the curvature at $\phi_0$ that is approximatively
given by $1/a$. For example, with the set parameters chosen in
Sec.~\ref{s:simu} one finds $T_R \sim 200$ in model units.

In bulk semiconductor lasers the residence times are
generally much larger than $T_R$ (the switching time between the two
states). Typically, $T_R\sim 5-10 ns$ while residence times
may range between 1 and 100 $\mu s$. The noise correlation time can
be somehow tuned but the noise generator limit $\tau$ to be larger
of 100 $ns$ (8.8 Mhz is the  maximal bandwidth used in the
experiment, see Section \ref{s:exp} below). Moreover, the
frequency of the relaxation oscillations of the total power is
typically above 1 GHz, and its damping occurs on a $ns$ time
scale, in accordance with the assumptions made in the theoretical
analysis.

At least to a first approximation, we can thus consider the limit
of large $\tau$. This justifies a further reduction of the problem
to a kinetic description which amounts to neglect intrawell
motion and reduce to a rate model describing the statistical
transitions in terms of transition rates. If we consider $\tau$ as
a time scale of the external driving we can follow the terminology
of Ref.~\cite{talk} and refer to this situation as the
``semiadiabatic" limit of Eq.~(\ref{langmu}). We will now consider
separately two limit cases.

\subsubsection{Slow barrier, rare hops: $T_R \ll \tau \ll T_\pm$}

This corresponds to the situation in which spontaneous emission
noise is very weak. In this case the residence time is
basically the shortest escape time, which in turn correspond to the
lowest value of the barrier (the noise is approximatively constant
in the current range considered henceforth). For the case of
interest,  $\delta < 2\varepsilon \sigma$ we can use (\ref{fbar}) to
infer that the minimal values of $\Delta U_\pm$ should be attained
for $\delta J \propto \mp\sqrt{D_J}$ respectively. This yields
\begin{equation}
T_\pm \;\simeq\;
T_s \exp\Big[-K \frac{2\varepsilon\sigma\pm\delta}{1+\sigma}
\,\frac{\sqrt{D_J}}{D_\phi} \Big]
\label{hopt}
\end{equation}
where $K$ is a suitable numerical constant. The ratio of
residence times $\eta$ is thus exponential in the noise
RMS:
\begin{equation}
\eta \;\equiv\; {T_+ \over T_-} \;=\; \exp\Big(-K \frac{2\delta}{1+\sigma}
{\sqrt{D_J} \over D_\phi} \Big)
\label{ratio}
\end{equation}
Notice that $\delta$ controls the asymmetry level: if $\delta \ll
2\varepsilon\sigma$ the two residence times decrease at
approximatively the same rate. This prediction is verified in the
simulations reported below and also in the experiment.

\subsubsection{Slow barrier, frequent hops: $T_R \ll T_\pm \ll \tau$}

This corresponds to the adiabatic limit in which the time scale of the external
driving is slower than the intrinsic dynamics of the system \cite{talk}.  The
results of this subsection are thus not of direct  relevance for interpreting
the experimental results reported here. However, we discuss also this regime
for completeness and to emphasize the differences with respect to the previous
case.

To a first approximation we can here treat current fluctuations
in a parametric way. The switch time will be the average of
escape times over the distribution of barrier fluctuations, i.e.
$\langle T_\pm \rangle_{\delta J}$.
Using again the approximation (\ref{fbar}),
since the variable $\delta J$ is Gaussian, using the identity
$\langle \exp{\beta z}\rangle= \exp(\beta^2\langle z^2\rangle /2)$
one gets \cite{m96}
\begin{equation}
T_\pm \;\simeq\; T_s \exp\Big[
{2(\delta\pm 2\varepsilon\sigma)^2 \over (1+\sigma)^2 D_\phi^2} \,
D_J \Big] .
\end{equation}
This means that in this limit the residence times increase
exponentially with the noise variance (i.e. exponentially with the
square of RMS). In other words, the fluctuations always increase the
effective barriers albeit asymmetrically.

\subsection{Comparison with the rate equations in the presence of current noise}

The above analysis shows that the effect of multiplicative noise may
strongly depend on the actual parameters of the model, in particular
of the ratio between $\varepsilon$ and $\delta$. Accordingly, the
simmetry-breaking effects can vary considerably depending on the
parameter $\delta$, see Eq.~(\ref{ratio}). To further clarify this
dependence and as a check of the prediction, we performed some
simulations of the rate equations (\ref{rateq1}-\ref{rateq3})
together with (\ref{ou}). We investigated the effect of changing the
value of the cross-saturation coefficient $c$, keeping  the other
parameters fixed the same used in Section \ref{s:simu}. The DC value
of the current $J$ has been again choosen empirically to yield an
almost symmetric distribution $P(\phi)$ for $D_J=0$.

\begin{figure}[h]
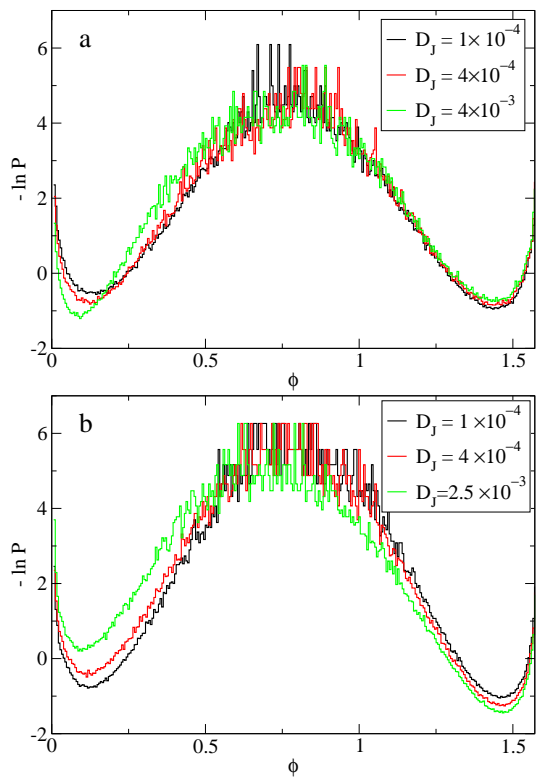

\begin{center}
\includegraphics[clip,width=7cm]{fig2a.eps}
\includegraphics[clip,width=7cm]{fig2b.eps}
\caption{
(Color online)
Simulation of the rate equation with current fluctuations:
$-\ln P$ for different amplitudes of imposed noise
$D_J$; for $c=1.1$, $J=1.197$, $D_{sp}=7 \times 10^{-6}$ (a)
and $c=1.3$, $J=1.194$, $D_{sp}=1.5 \times 10^{-5}$ (b).}
\label{f:lnp}
\end{center}
\end{figure}

We set $\tau=4 \times 10^{4}$ that correspond
roughly to the experimental values. In Fig.~\ref{f:lnp} we
plot $-\ln P$ for $c=1.1$ and $c=1.3$ corresponding to $\delta=0.05$ and
$\delta=0.15$ respectively. In the second case  the
multiplicative effect of the noise is stronger and leads to  a sizeable
distortion of the curve while for small asymmetries  the curves basically
mantain their symmetry.

\begin{figure}[h]
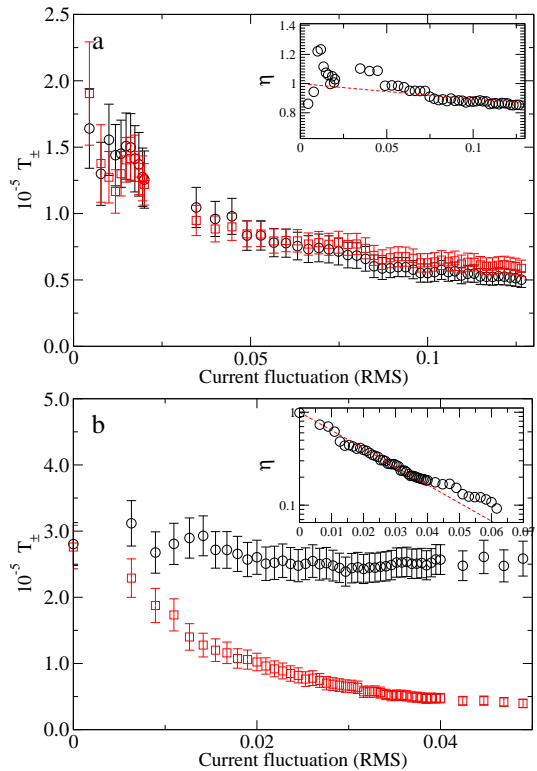

\begin{center}
\includegraphics[clip,width=7cm]{fig3a.eps}
\includegraphics[clip,width=7cm]{fig3b.eps}
\caption{(Color online)
Simulation of the rate equation with current fluctuations:
Residence times $T_+$ (squares) and $T_-$ (circles) 
for $c=1.1$ (a) and $c=1.3$
(b). The parameters are the same as in Fig.~\ref{f:lnp}.
The insets report the ratio of the two times and the dashed line
is an exponential fit, see expression (\ref{ratio}).}
\label{f:tk2}
\end{center}
\end{figure}

The same type of behavior can be observed in the dependence of the
residence times on the current fluctuations.
Fig.~\ref{f:tk2}a illustrates how for $c=1.1$ ($\delta=0.05$) the
residence times decrease at approximatively the same rate
upon increasing current fluctuations and the ratio $\eta$ remains
very close to 1. On the contrary, upon increasing the
cross-saturation coefficient to $c=1.3$ ($\delta=0.15$) a sizeable
symmetry breaking (Fig.~\ref{f:tk2}b): one of the two times remains
almost constant while the other decreases. This is in agreement with
formulae (\ref{hopt}) and (\ref{ratio}). Indeed, the ratio between
the decay rates of $\eta$ obtained from the fits (insets of
Fig.~\ref{f:tk2}) is about 40, which is roughly a factor 2 of the
corresponding value as computed from (\ref{ratio}) with the
simulation parameters at hand.

\section{Experimental Results}
\label{s:exp}

\subsection{Experimental set-up}

The experimental setup is similar to the one described in
\cite{IEEE,AppB}. Several edge-emitting semiconductor laser were tested:
two Hitachi Hlp 1400 lasing at 840 nm and three Sharp LT021 MD lasing at 780
nm.  These lasers are GaAlAs double-hetero-structure with a bulk active region
and cleaved, uncoated facets. The wavelength separation between consecutive
longitudinal-modes is around 0.3~nm, and the laser emission occurs in a
single-transverse mode. The laser package temperature is stabilized up to $\rm
0.01^{o}C$ and the laser current is controlled with a very stable (up to $\rm
1~\mu A$) power supply.  The laser is optically isolated from the rest of the
set-up by means of an optical diode that avoid spurious back reflection. The
total laser emission is detected by an Avalanche Photodiode detector (APD)
(DC-1.5~GHz bandwidth) while the time-averaged optical spectrum is measured by
an Agilent 86140B spectrum analyzer. Individual longitudinal mode detection is
obtained by sending part of the laser output into a monochromator (resolution
$0.5~\AA$) with two independent output slits. The monochromator can be set in
order to have at the two exits two different longitudinal modes emitted by the
laser. Each longitudinal mode intensity is then monitored by two APD detectors
(DC-1.5~GHz bandwidth) placed beyond the slits. The outputs from these
detectors are monitored on a LeCroy 7200A (500 MHz analogue bandwidth, 1~GS/s).
The power spectra of the signals can be monitored using an Agilent~E4403B
spectrum analyzer.

\subsection{Longitudinal mode-hopping}

Two control parameters can be directly varied in this system: the
pumping current $J$ and the temperature of the laser substrate
$T_{sub}$. As described in \cite{IEEE,AppB}, the optical spectra of the
lasers tested show a single longitudinal mode emission for most of the
parameter values ($J$, $T_{sub}$). However,  there exist small regions
in the parameter space where the lasers are bistable and exhibit
mode-hopping between two longitudinal modes. A basic explanation of the
destabilization of the leading longitudinal mode is that, fixing the
$T_{sub}$ and increasing $J$ the wavelength of the cavity resonance
increases due to Joule heating of the semiconductor medium. The gain
curve peak shifts towards longer wavelength as well, but at a larger
rate. Eventually, the dominating longitudinal mode looses its stability
in favour of a longer wavelength one,  which has become more resonant
with the gain peak. The same happens when, for fixed $J$, $T_{sub}$ is
increased. This type of transition among lasing modes is well-known
\cite{petermann} and occurs sharply in the parameter space.

Observation of this regime of stochastic mode-hopping has been
reported in several papers \cite{Ohtsu,Ohtsu2,IEEE}. Its most
relevant features are: (i) the total emitted intensity remains
almost constant though each longitudinal mode is switching on and
off. In other words, the anti-correlation of the two modal
intensities is very strong (better than $-0.95$ in our case); (ii)
the distribution of residence times in a mode, defined as the
interval between a switching-on and a switching-off event, is
Poissonian; (iii) a sweep of the pump current across this
parameters' region, reveals the existence of an hysteresis cycle for
the modal emission which is a signature of bistability between the
two longitudinal modes. It is thus clear that these features are
straightforwardly explained by the theoretical analysis presented
above.

\begin{figure}
\includegraphics[width=8cm]{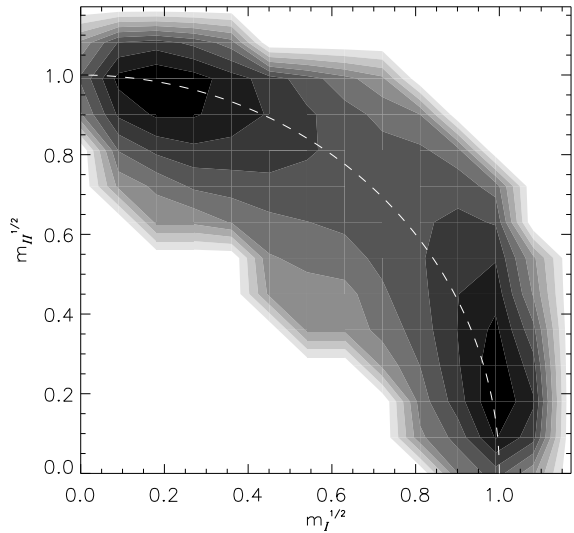}
\caption{ Experimental probability distribution  of the normalized
modal amplitudes (see text). The histogram is represented in
logarithmic scale with a grey scale code from white (low
probability) to black (high probability); $J=95.9 mA$,
$T_{sub}=20.4^oC$.} \label{fig2expe}
\end{figure}

The study of the fluctuations of the laser emission can be carried
out  by monitoring the modal intensities normalized to the total
intensity output: $m_I=i_I/i_{tot}$ and $m_{II}=i_{II}/i_{tot}$. The
mode labeled with $I$ is the one with larger wavelength. These are
related to $\rho_\pm$ defined in the theoretical analysis,  by
$m_{I,II}=\rho_\pm^2/r_0^2$. Then, it is useful to represent the
state of the system in the phase space
(${m_I}^{1/2},{m_{II}}^{1/2}$).  Hence, the variable $\phi$ used in
the theory is simply the angle that this vector forms with the
horizontal axis. The strong anticorrelation  of the modal
intensities implies that the modulus of this vector is almost
constant and equal to one. In Fig.~\ref{fig2expe} we plot the the
probability distribution function in the space
(${m_I}^{1/2},{m_{II}}^{1/2}$). The maxima of the distribution
correspond to the two stable modes involved in the hopping. It is
evident that the relevant dynamics occur on a portion of the unit
circle as assumed in the simplified theory.

\begin{figure}
\includegraphics[clip,width=8cm]{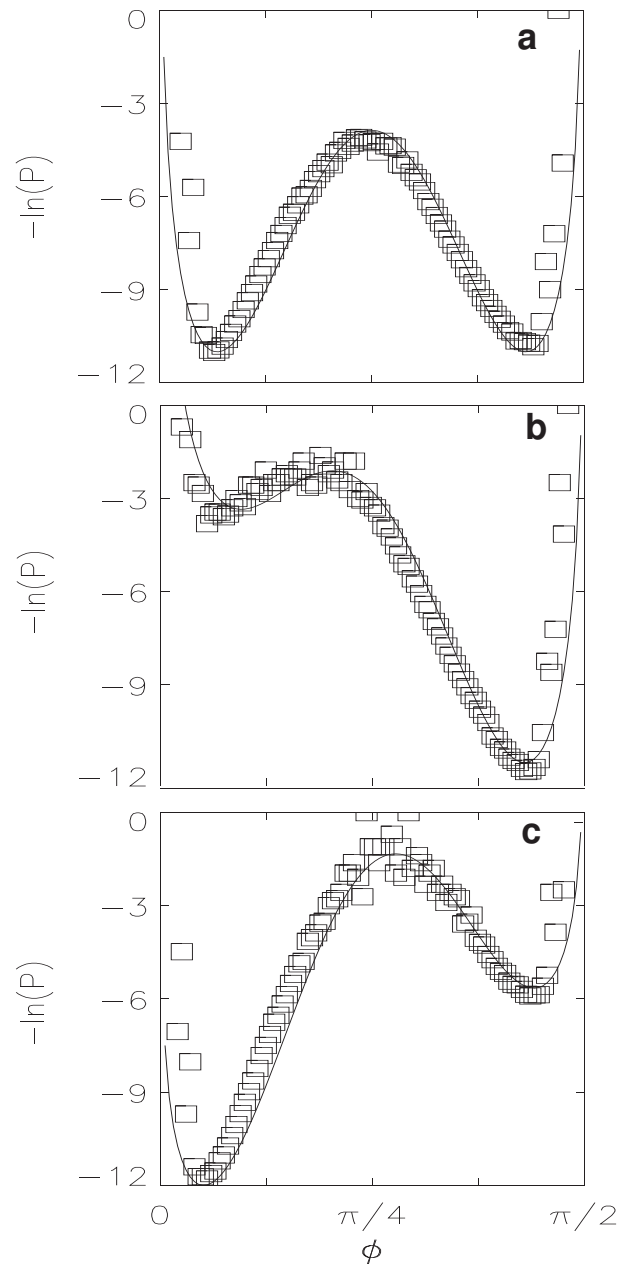}
\caption{ The logarithm of the experimental probability distribution
function  of the variable $\phi$ (squares) for three different
values of the current, $T_{sub}=20.4^oC$. The line is the fitting
with the function given in the text. (a) $J=95.9 mA$, fitting
parameters $A=108.5, B=0, D=4.6, C=-10.6$; (b) $J=95.1 mA$, fitting
parameters $A=85, B=-18.7, D=4.9, C=-8.1$; (c) $J=96.6 mA$, fitting
parameters $A=90, B=13.5, D=2.5, C=-7.2$.} \label{fig3expe}
\end{figure}

The probability distribution functions of the angular variable
$\phi$ as obtained from the experiment are reported in
Fig.~\ref{fig3expe} for different values of the pumping current. For
every $\phi$ we have included in the histograms all the points along
the radial direction. Two peaks located around $\phi = 0.2 $ and
$\phi = 1.4 $ are found, corresponding to the activation of the mode
$I$ and $II$ respectively. In Fig.~\ref{fig3expe}a we plot the
distribution of $\phi$ for the case of Fig.~\ref{fig2expe}, where
the modal emission is symmetric, $J\simeq J_s$. The measurements
show that upon changing the pumping current an asymmetric situation
settles down,  where the laser emission on one of the two modes is
favored. In particular, an decrease (resp. increase) of $J$ with
respect the value of Fig. ~\ref{fig2expe}, enhances the stability of
the mode labeled by $II$ (resp. $I$), see Fig. \ref{fig3expe}b,c
respectively.  In all cases, the logarithm of the distribution can
be fitted with the function (see Eqs.~(\ref{uphi}) and (\ref{ps}))
$-\frac{A}{16}\cos 4\phi - \frac{B}{4}\cos 2\phi
 - D \ln\sin 2\phi + C$.

\begin{figure}
\includegraphics[clip,width=8cm]{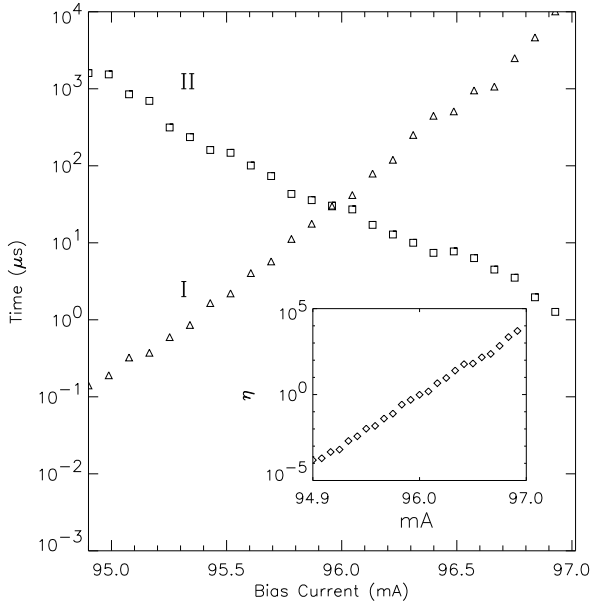}
\caption{Average residence times for mode  $I$
(triangles) and $II$ (squares) as a function of pumping current,
$T_{sub}$=$20.4^oC$. Inset: the ratio $T_{I}/T_{II}$.}
\label{fig1expe}
\end{figure}

The dependence of the average residence times, $T_I$ and
$T_{II}$, on the pumping current is shown in Fig.~\ref{fig1expe}. As
predicted by formula (\ref{tpm}), both times depend exponentially on
$J$ and the two curves have different absolute values of the slopes
(remember the remark at the end of Sec.~\ref{s:bist}). Incidentally,
this implies that the quantity $(T_{I}+T_{II})/2$ increases with the
current, meaning that hops become more and more rare. Moreover, the
ratio $\eta = T_{I}/T_{II}$ increases  exponentially with the
current as shown in the inset of Fig.~\ref{fig1expe}.  This
behaviour is also consistent with formula (\ref{tpm}).

\subsection{Effect of noise addition onto the pumping current}

In a semiconductor laser driven by a constant ($DC$) current, the
main source of noise is the spontaneous emission within the
semiconductor medium. This noise source cannot be varied directly
through the control parameters. Fluctuations can be added into the
system externally by adding electrical noise on the pumping current.
This signal employed in our setup has zero-mean, a bandwidth ranging
from 100 Hz to 8.8 MHz and it is AC-coupled to the laser bias
current.

\begin{figure}
\includegraphics[clip,width=7cm]{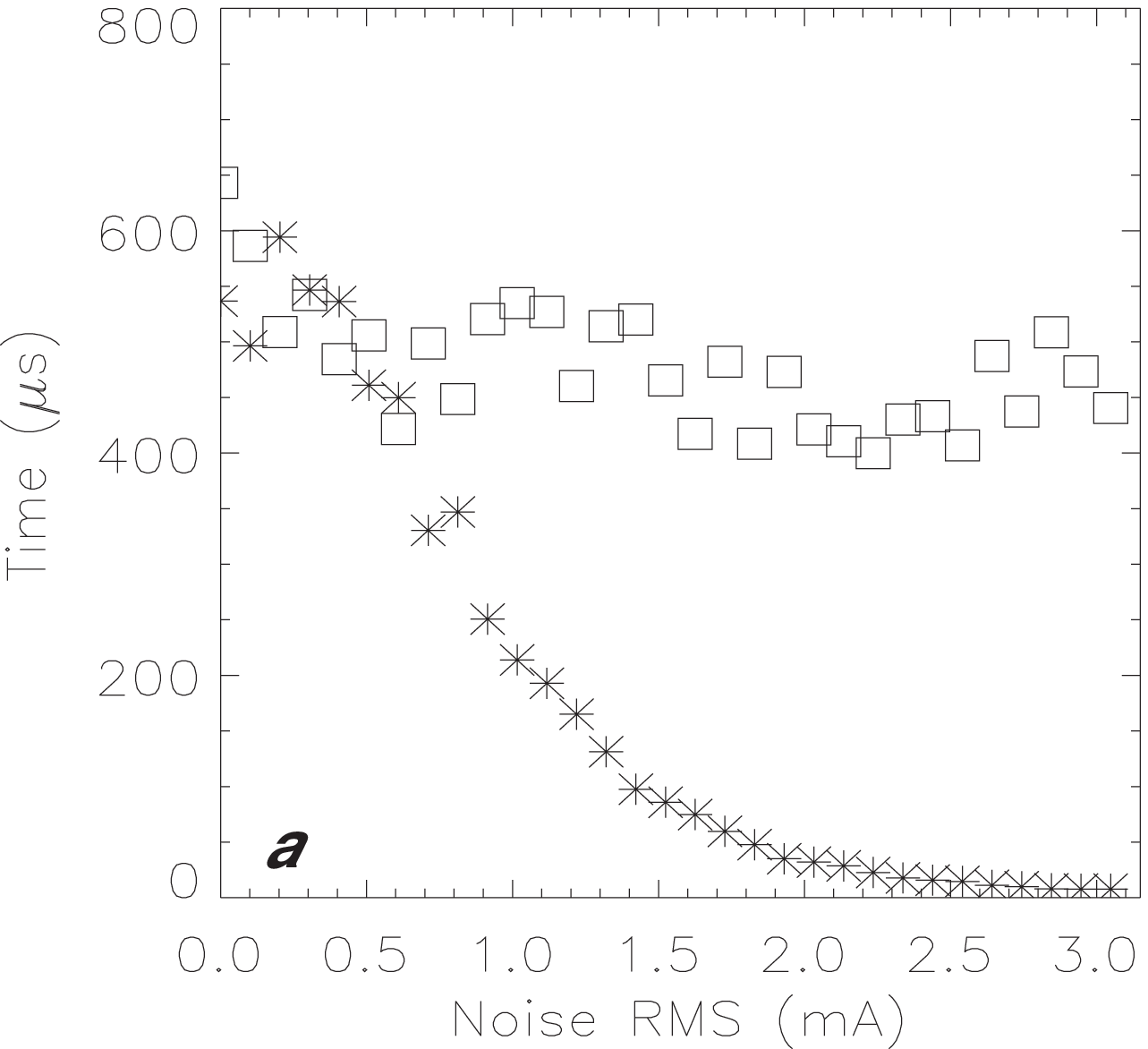}
\includegraphics[clip,width=8cm]{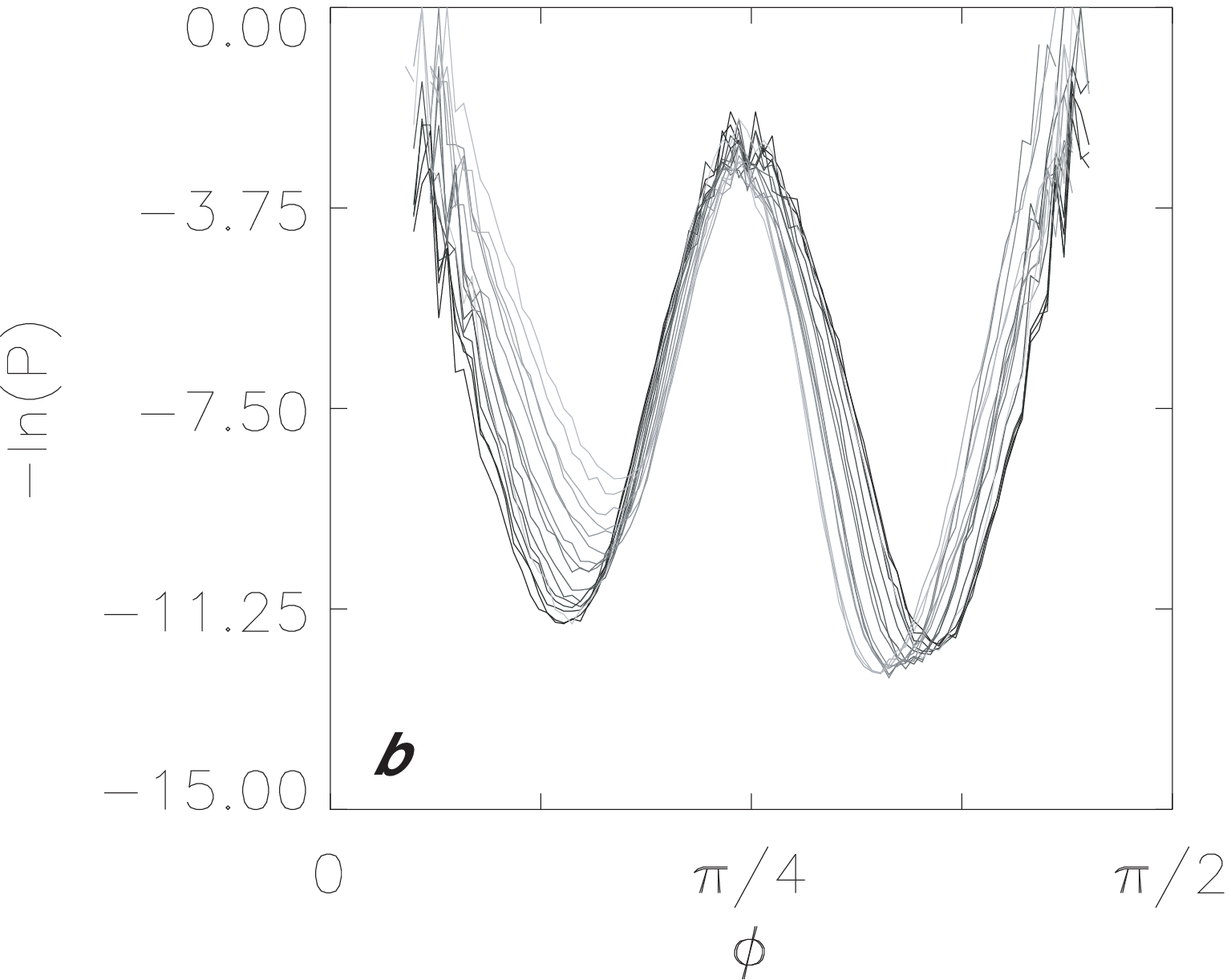}
\caption{Effect of external noise added on the pump current. The
measurements are performed on a laser sample different from the one
used for Figs. \ref{fig2expe},\ref{fig3expe}, \ref{fig1expe},
requiring slightely modified values for the control parameters to
obtain the same behaviors. (a) The average residence times
$T_{I}$ (stars) and $T_{II}$ (squares) as a function of the RMS level
of the noise added to the pumping current, $T_{sub}=20.5^oC$,
$J$=92.0~mA. (b) The corresponding logarithm of probability
distribution function of the variable $\phi$ for different levels of
current noise (noise increases from dark trace to grey trace).}
\label{fig4expe}
\end{figure}

\begin{figure}
\includegraphics[clip,width=8cm]{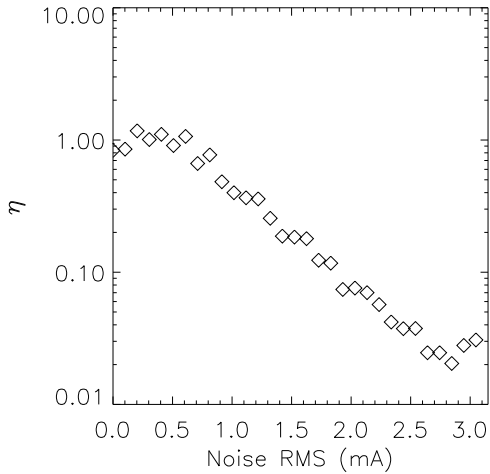}
\caption{The ratio of residence times $\eta=T_{I}/T_{II}$
as a function of the RMS level of the noise added to the pumping
current (parameters are the same of Fig. \ref{fig4expe}).}
\label{fig5expe}
\end{figure}

In Fig.~\ref{fig4expe}a we show how the average residence times
$T_{I,II}$ change upon increasing the current noise level around the
static value $J_s$.  The times are affected in a strongly asymmetric
way: while $T_{II}$ is almost unchanged, $T_I$ decreases
exponentially. In Fig.~\ref{fig4expe}b we plot the probability
distribution function of the angular variable $\phi$ for different
values of the pump noise. The experimental evidence confirms that the
asymmetry of the probability distribution increases upon increasing
the noise level. It is interesting to remark that an increase of
noise level is equivalent, from the point of view of the residence
times, to a decrease of the $DC$ pumping current. This experimental
result confirms that the imposed noise does affect the dynamics in a
multiplicative way. The qualitative agreement between the
experimental data and the simulations reported in Figs.~\ref{f:lnp}b
and \ref{f:tk2}b supports the validity of our modelling. The
exponential dependence of the ratio $\eta$ on the RMS of the
fluctuation predicted by the analysis of the reduced model,
Eq.~(\ref{ratio}), is indeed observed in the experiment (see
Fig.~\ref{fig5expe}). This is another indication that confirms that
the kinetic description suffices to capture the mode-hopping
dependence on noise level.

\section{Conclusions}

In this paper we have explored experimentally and theoretically the
effects of external current noise on the mode-hopping dynamics of a
bistable semiconductor laser. We have shown that the residence times
of each mode are strongly affected, bearing the typical signatures
of multiplicative noise. We have investigate a theoretical model
based on a rate-equation description, where the bias current enters
parametrically into the evolution of the modal amplitudes, hence the
multiplicative character of its fluctuations. Numerical simulations
of the rate equations are in good qualitative agreement with the
experimental observations. Moreover, the reduction of this model to
a 1D Langevin equation describing activated  escape over a
fluctuating barrier has allowed us to draw some  predictions (e.g.
the dependence of residence times on noise strength) and to
better understand the role of the physical parameters. Depending on
their values, the fluctuating part of the effective potential may
have a different parity  with respect to the static one. This
explains the observation that imposed fluctuations have different
effects on the hopping rates (symmetry breaking).

The reduced model and many of the results presented here could be
useful to describe also polarization switching in VCSELs driven by
external noise. Indeed, experimental data show strong similarities
between this phenomenon \cite{gianni3,gianni4} and the longitudinal
mode dynamics. On the theoretical side, this analogy is supported by
the fact that the polarization dynamics in VCSELs is described
by models that are very similar to the one discussed here
\cite{will,will2,belgi}. The inclusion of current noise effects
along the same line above reported should be feasible once the
typical time scales and parameters for such a class of lasers
are evaluated. This work, that we plan to undertake in the
future, will provide a common theoretical basis to the stochastic
dynamics of bistable semiconductor lasers.

\acknowledgments

G. G. acknowledges partial support by INLN. S. B. acknowledges
partial financial support from ESF, project STOCHDYN 292. F.P. and
M.G. acknowledge Gabriel Mindlin and St\'{e}phane Barland for the
useful discussions.


\begin{thebibliography}{99}
\bibitem{note} Actually this statement is not correct when $\delta \simeq
2\varepsilon\sigma$. As seen from Eq.~(\ref{barris}), in this case
$\Delta U_-$ is hardly affected by a change in $J$ and the change of
noise amplitude becomes relevant, Since the parameters are independent
we restrict to the generic case in which the above condition is not
fullfilled.
\end{thebibliography}
\end{document}